\documentclass[useAMS,usenatbib]{mn2e}
\usepackage{graphics}
\usepackage{epsfig}

\def\lta{\mathrel{\spose{\lower 3pt\hbox{$\mathchar"218$}} \raise
2.0pt\hbox{$\mathchar"13C$}}} \def\gta{\mathrel{\spose{\lower
3pt\hbox{$\mathchar"218$}} \raise 2.0pt\hbox{$\mathchar"13E$}}}
  
\title{Correlation Functions for Extended Mass Galaxy Clusters}
\author[]{Naseer Iqbal\thanks{E-mail: iqbal@iucaa.ernet.in}, Naveel Ahmad, Mubashir Hamid and Tabasum Masood\\
Department of Physics, University of Kashmir, Srinagar-190006} 
\begin{document}
\date{}
\pagerange{\pageref{firstpage}--\pageref{lastpage}} \pubyear{2012}
\maketitle

\label{firstpage}

\begin{abstract}
The phenomenon of clustering of galaxies on the basis of correlation functions in an expanding universe is studied by using equation of state, taking gravitational interaction between galaxies of extended nature into consideration. The partial differential equation for the extended mass structures of a two-point correlation function developed earlier by \citet{Iq06} is studied on the basis of assigned boundary conditions. The solution for the correlation function for extended structures satisfies the basic boundary conditions, which seem to be sufficient for understanding the phenomena and provides a new insight into the gravitational clustering problem for extended mass structures.
\end{abstract}
\section{Introduction}
The universe is dominated by matter in gravitational interaction into the spatial configuration consisting of galaxies, group of galaxies, super-clusters and even larger structures \citet{Sh67} and \citet{Ma90}. One of the standard ways in which structures in the universe were formed (i.e; the departure from randomness and homogeneity) is described by means of correlation functions. From observation the galaxy-to-galaxy correlation function $\xi_{gal}$ is known to scale as $\xi_{gal}=r^{-\gamma}$, where the exponent $\gamma$ ranges from 1.6 to 1.8. The calculation of $\gamma$ tacitly assumes the universe to have evolved to the stage, where the initial primordial matter has been formed into the observed galaxies and these galaxies are coupled to the expansion of the universe. We take our clue by analogy with the well established theory of interacting gases \citet{Go85}, where the relative arrangements of the atoms and molecules (galaxies in our case) are accurately described in terms of the principles and methods of thermodynamics. We apply the same techniques to a system made up of many galaxies in gravitational interaction, subject to fluctuations emerging from the intrinsic properties of the gravitational interaction. A number of theories of the cosmological many body problem have been developed mainly from the thermodynamic point of view \citet{Sa84}, \citet{Sa00} and \citet{Iq06}. The theory makes use of the equations of state along with the correlation functions for the extended mass structures for the development of a semi-analytical model. This can even be done by solving a system of Liouville's equation or BBGKY-hierarchy equations and have been discussed by  \citet{Pe80} and \citet{Sa84}. But BBGKY-hierarchy equations are too complicated to handle for higher order correlation functions. We extract the possible information about lowest order i.e; two-point correlation function for galaxies with real extended mass structures clustering gravitationally in an expanding universe. Although, the same information about $\xi_{2}$ have earlier been developed by \citet{Iq06}, but here we have studied and gather the related information about $\xi_{2}$ for extended mass structures. It is important to note that galaxies with point mass consideration is only an approximation. In fact, galaxies have real extended structures, where dark matter especially is having an important contribution.
\section{Two-point correlation function for extended mass structures}
 The general pair of equations of state for internal energy ($U_{e}$) and  pressure ($P_{e}$) for extended mass structures is given as;
\begin{equation}
           U_{e}=\frac{3}{2}NT-\frac{2\pi Gm^{2}N^{2}}{V}\int\xi(\bar{n},T,r)\left(1+\frac{\epsilon^{2}}{r^{2}}\right)^\frac{-1}{2}\frac{dV}{4\pi r}
\end{equation}
\begin{equation}
           P_{e}=\frac{NT}{V}-\frac{2\pi Gm^{2}N^{2}}{3V^{2}}\int\xi(\bar{n},T,r)\left(1+\frac{\epsilon^{2}}{r^{2}}\right)^\frac{-3}{2}\frac{dV}{4\pi r}
\end{equation} 
Equations(1) and (2) represent the equations of state \citet{Hi56} and \citet{Iq06}. The measuring correlation parameter ($b_{e}$) for extended mass structures is given by
\begin{equation}
         b_{e}=\frac{2\pi Gm^{2}N}{3VT}\int\xi(\bar{n},T,r)\left(1+\frac{\epsilon^{2}}{r^{2}}\right)^\frac{-1}{2}\frac{dV}{4\pi r}
\end{equation}
    Here $\bar{n}=\frac{N}{V}$ is the average number density of extended mass  particles (galaxies) with $\epsilon$ as softening parameter and is taken between 0.01 to 0.05 (in the units of total radius), T is the temperature, V the volume, G is the gravitational constant, $\xi(\bar{n},T,r)$ is the two-point correlation function for extended mass structures and 'r' the intergalactic  distance. It should be noted that these expressions assume a large volume V  for their validity. \par
The Maxwell's thermodynamic equation relating internal energy ($U_{e}$) and pressure ($P_{e}$) for extended mass structures is given by  
\begin{equation}
   \left(\frac{\partial U_{e}}{\partial V}\right)_{T,N}=T\left(\frac{\partial P_{e}}{\partial T}\right)_{V,N}-P_{e}
\end{equation}
From equations (1) and (2), we have
\begin{eqnarray}
\left(\frac{\partial U_{e}}{\partial V}\right)_{T,N}&=&\frac{-2\pi G m^{2} N^{2}}{V}\frac{\xi(\bar{n},T,r)}{4\pi r}\left({1+\frac{\epsilon^{2}}{r^{2}}}\right)^{\frac{-1}{2}}\nonumber\\
&&+\frac{2\pi G m^{2} N^{2}}{V^{2}}\nonumber\\
&&\int\frac{\xi(\bar{n},T,r)}{4\pi r}\left({1+\frac{\epsilon^{2}}{r^{2}}}\right)^{\frac{-1}{2}}dV
\end{eqnarray}
and\\
\begin{eqnarray}
\left(\frac{\partial P_{e}}{\partial T}\right)_{V,N}&=&\frac{N}{V}-\left(\frac{2\pi G m^{2} N^{2}}{3V^{2}}\right)\nonumber\\
&&\int\left(1+\frac{\epsilon^{2}}{r^{2}}\right)^{\frac{-3}{2}}\frac{\partial\xi(\bar{n},T,r)}{\partial{T}}\frac{dV}{4 \pi r}
\end{eqnarray}\par
Using the equations (5) and (6) in equation (4), and then differentiating the required equation with respect to V, the two-point correlation differential equation for extended mass structure takes the form as \citet{Iq06};
\begin{equation}
3\bar{n}\frac{\partial\xi_2}{\partial{\bar{n}}}+T\left(\frac{r^2}{r^2+\epsilon^2}\right)\frac{\partial\xi_2}{\partial T}-r\frac{\partial\xi_2}{\partial r}=0
\end{equation}\par
Here $\xi_{2}$ depends on the variables $\bar{n}$, T, and r. In order to look for the possible solution of this equation, we write $\xi_{2}$ as product of three variables defined by;\\
$\Theta(\bar{n})$, $\Gamma(T)$, and R(r).\par
\begin{equation}
\xi_{2}(\bar{n},T,r)=\Theta(\bar{n})\Gamma(T)R(r)
\end{equation}
Calculating the required terms from equation (8), equation(7) gets modified to\\
\begin{equation}
\frac{3\bar{n}}{\Theta}\frac{d\Theta(\bar{n})}{d\bar{n}}=\frac{r}{R}\frac{dR(r)}{dr}-\frac{T}{\Gamma}\frac{r^2}{\epsilon^2+r^2}\frac{d\Gamma(T)}{dT}
\end{equation}

As both sides of equation (9) are functions of different variables, so equation (9) will be true only, if both sides can be equal to the same constant say 'Z'.
\begin{equation}
\frac{3\bar{n}}{\Theta}\frac{d\Theta(\bar{n})}{d\bar{n}}=Z
\end{equation}
and
\begin{equation}
\frac{r}{R}\frac{dR(r)}{dr}-\frac{T}{\Gamma}\frac{r^2}{\epsilon^2+r^2}\frac{d\Gamma(T)}{dT}=Z
\end{equation}
Equation (10) leads to the solution as;
\begin{equation}
\Theta(\bar n)=C_{1}(\bar n)^\frac{Z}{3}
\end{equation}
with $C_{1}$ as constant.\\
Similarly, in order to find the solution of the equation (11), we can proceed as;
\begin{equation}
\frac{\epsilon^2+r^2}{r^2}\left(\frac{r}{R}\frac{dR}{dr}-Z\right)=\frac{T}{\Gamma}\frac{d\Gamma}{dT}
 \end{equation} 
Equation (13) can be correct only,
if both sides of it are equal to the same constant say $'Z_N'$,
\begin{equation}
\frac{\epsilon^2+r^2}{r^2}\left(\frac{r}{R}\frac{dR}{dr}-Z\right)=Z_N
\end{equation}
\begin{equation}
\frac{T}{\Gamma}\frac{d\Gamma}{dT}=Z_N
 \end{equation}
The possible solutions of equations (14) and (15) are;
\begin{equation}
R(r)=C_{2}r^{Z}(\epsilon^2+r^2)^\frac{Z_N}{2}
\end{equation}
\begin{equation}
\Gamma(T)=C_{3}T^{Z_N}
 \end{equation}
with $C_{2}$ and $C_{3}$ as constants.\\
With the help of the above possible solutions, the two point correlation function defined by equation (8) is 
\begin{equation}
\xi_{2}(\bar{n},T,r)=C (\bar n)^\frac{Z}{3}T^{Z_N}r^{Z}(\epsilon^2+r^2)^\frac{Z_N}{2}
\end{equation}
where $C=C_{1}C_{2}C_{3}$
\section{ Description of Correlation Function and Correlation Energy for Extended Mass Structures}
The solution defined by equation (18) can have variety of forms, depending upon the parameters Z and $Z_{N}$, but we are interested in such a solution that is physically valid and satisfies earlier results also. A set of boundary conditions assumed here are as;
\begin{itemize}
\item The gravitational clustering of galaxies in a homogeneous universe requires $\xi_{2}$ to have a positive value, which obviously depends upon the limiting values of  $\bar{n}$, T, r and $\epsilon$.
\item When $\bar{n}T^{-3}$ is very large, the two-particle correlation function $\xi_{2}$ will increase and measuring correlation energy $b_{e}$ will also increase, and vice versa.
\end{itemize}
So depending upon the boundary conditions of $\xi_{2}$ as per the requirement, we choose the values of Z and $Z_{N}$. The substitution of equation (18) in equation (3) gives;
\begin{equation}
b_{e}=\frac{2\pi Gm^{2}\bar{n}}{3T}C\left(\frac{3N}{4\pi}\right)^{\frac{Z}{3}}T^{Z_{N}}\int_{0}^R r^{2}{(r^{2}+\epsilon^{2})^\frac{Z_{N}-1}{2}}dr
\end{equation}
It can be seen, that for different values of $Z_{N}$, the integral will have different values. We fix a set of values for $Z_{N}$ starting from unity (1) and check the correlation energy for other values also like 0, -1, -2 etc. The following cases are discussed here for illustration.\\

For $Z_{N}=1$, equation (19) leads to;
 \begin{equation}
b_{e}=\frac{2\pi Gm^{2}\bar{n}}{3}C\left(\frac{3N}{4\pi}\right)^{\frac{Z}{3}}\frac{R^{3}}{3}
\end{equation}
Equation (20) relates the variation of $b_{e}$ (the measuring correlation parameter for extended mass structures) with the size of cluster. For a given cluster cell with more dimensions of R, we can study the clustering rate without involving the thermodynamic quantities. However, there is need to show the temperature dependence of $b_{e}$ also, as we have assumed the system in quasi-equilibrium state. This is achieved by testing for other values of $Z_{N}$. It is interesting to note here that the validity of $Z_{N}$ is based on $Z_{N} \le 1$.\\

For $Z_{N}=0$, equation (19) becomes as;
\begin{equation}
b_{e}=\frac{2\pi Gm^{2}\bar{n}}{3T}C\left(\frac{3N}{4\pi}\right)^{\frac{Z}{3}}\int_{0}^R \frac{r^{2}}{(r^{2}+\epsilon^{2})^\frac{1}{2}}dr
\end{equation}
\begin{eqnarray}
b_{e}&=&\frac{2\pi Gm^{2}\bar{n}}{3}C\left(\frac{3N}{4\pi}\right)^{\frac{Z}{3}}T^{-1}\frac{1}{2}\nonumber\\
&&\left(R\sqrt{{\epsilon^{2}}+{R^{2}}}-\epsilon^{2}log\left|\frac{R}{\epsilon}+\sqrt{1+\frac{R^{2}}{\epsilon^{2}}}\right|\right)
\end{eqnarray}\par
For $Z_{N}=-1$, we have;
\begin{equation}
b_{e}=\frac{2\pi Gm^{2}\bar{n}}{3T}C\left(\frac{3N}{4\pi}\right)^{\frac{Z}{3}}T^{-1}\int_{0}^R \frac{r^{2}}{(r^{2}+\epsilon^{2})}dr
\end{equation}
\begin{equation}
b_{e}=\frac{2\pi Gm^{2}\bar{n}}{3}C\left(\frac{3N}{4\pi}\right)^{\frac{Z}{3}}T^{-2}\left(R-\epsilon tan^{-1}\left(\frac{R}{\epsilon}\right)\right)
\end{equation}\par
Finally for $Z_{N}=-2$, we have;
\begin{equation}
b_{e}=\frac{2\pi Gm^{2}\bar{n}}{3T}C\left(\frac{3N}{4\pi}\right)^{\frac{Z}{3}}T^{-2}\int_{0}^R \frac{r^{2}}{(r^{2}+\epsilon^{2})^\frac{3}{2}}dr
\end{equation}
\begin{eqnarray}
b_{e}&=&\frac{2\pi Gm^{2}\bar{n}}{3}C\left(\frac{3N}{4\pi}\right)^{\frac{Z}{3}}T^{-3}\nonumber\\
&&\left(log\left|\frac{R}{\epsilon}+\sqrt{1+\frac{R^{2}}{\epsilon^{2}}}\right|-\frac{R}{\sqrt{R^{2}+\epsilon^{2}}}\right)
\end{eqnarray}\par 
The equations (22), (24), and (26) are in good agreement with a set of boundary conditions. The correlation is maximum, if galaxies are treated as point mass objects ($\epsilon =0$) and goes on decreasing as softening parameter ($\epsilon$) goes on increasing, thus clustering decreases. It has been found that the dispersion in the radial velocity could be up to 1000 $Km s^{-1}$(e.g coma cluster) and is sufficient to throw galaxies in the surrounding voids. The cause for decrease of correlation for galaxies with extended structures, may be due to dispersion in radial velocity. The precise nature of such dependence between $b_{e}$ and the increase in radial velocity can be treated  as one of the important problem to study in contact to the correlation function studies for galaxy clusters. Equation (26) is important in the sense that it clarifies the earlier result as well as justifies the significance of $b_{e}$ and $b$. In the earlier work of \citet{Sa84} and \citet{Iq06}, it has been understood that $b$ has a specific dependence on the combination $\bar{n}T^{-3}$. Here in our study, it is also true that $b_{e}$ has also the specific dependence on $\bar{n}T^{-3}$, which means that the clustering takes place at moderate level. Z may have always positive values because of the reason that for large values of N, $b_{e}$ increases.  
\section{Discussion}
The two-point correlation function $\xi_{2}$ for point mass galaxies is the lowest possible tool for understanding the phenomena of galaxy clusters. In this paper, an attempt has been made to extend the theory of correlation function from point mass to extended mass structures by developing  a new set of equations on the basis of assigned values of the constant $Z_{N}$ ($Z_{N}$=1, 0, -1, -2). The measuring correlation parameter $b_{e}$ measures the effectiveness of the clustering rate on the basis of the two point correlation function $\xi_{2}$ and the results obtained are in good agreement with the earlier work of \citet{Iq06}. One of the interesting feature in this paper is to note how the theoretical values of $Z_{N}$ coincides with the observed values of $\gamma$ (1.6-1.8) on the same pattern. The other interesting result ($b_{e}$ having specific dependence on the combination $\bar{n}T^{-3}$) for the extended structures confirms the earlier results for point mass approximation where the functional form of $b=b(\bar{n}T^{-3})$. The solution of equation (7) given by equation (18) can be used now as an alternative tool for studying the phenomena of galaxy clusters on the basis of assigned values of $Z_{N}$.

\label{lastpage}

\end{document}